\documentclass[aps,prc,preprint,showpacs,groupedaddress,floatfix,amsmath,amssymb]{revtex4-2}

\usepackage{color}
\usepackage{graphicx}
\usepackage{longtable}
\usepackage{float}

\usepackage{subfigure}
\usepackage{multirow}

\usepackage{makecell}
\usepackage{braket}
\usepackage{bigstrut}
\usepackage[T1]{fontenc}

\usepackage[
pdfstartview=FitH, CJKbookmarks=true, bookmarksnumbered=true,
bookmarksopen=true, colorlinks=false, pdfborder=001, linkcolor=blue,
anchorcolor=blue, citecolor=blue ]{hyperref}

\newcommand{\beq}{
\begin{equation}}
  \newcommand{\eeq}{
\end{equation}}

\begin{document}

\title {$\Delta l =1$ coupling of single-particle orbitals in octupole deformed
nuclei} 
\author{XuDong Wang}
\author{Bin Qi} \email{bqi@sdu.edu.cn}
\author{Shouyu Wang}
\author{Chen Liu}
\affiliation{Shandong Provincial Key Laboratory of Nuclear Science, Nuclear 
Energy Technology and Comprehensive Utilization, Weihai Frontier Innovation 
Institute of Nuclear Technology, School of Nuclear Science, Energy and Power 
Engineering, Shandong University, Shandong 250061, China} 

\affiliation{ Weihai Research Institute of Industrial Technology of Shandong 
University, Weihai 264209, China} 

\begin{abstract}
Conventionally, octupole deformation in nuclei has been attributed to 
strong $\Delta l=3$ couplings between opposite-parity single-particle 
orbitals. In this work, we demonstrate that the often-overlooked $\Delta 
l=1$ mode also plays an important role. Taking orbitals near the octupole 
magic number $N = 134$ as a benchmark, we systematically evaluate the 
$\Delta l = 1$ and $\Delta l = 3$ mixing ratios of the wave functions 
within the Nilsson model, interpreting the trends through matrix elements 
of the deformed potential. We introduce component-resolved single-particle 
octupole energy contributions, based on the Hellmann--Feynman relation, to 
quantify the contributions of each $(\Delta l,\Delta j)$ coupling. 
Furthermore, the impact of $\Delta l = 1$ coupling on the rotational 
structure is demonstrated via particle-rotor model calculations for 
$^{221}$Ra and $^{223}$Th. Our work suggests that $\Delta l=1$ and $\Delta 
l=3$ octupole couplings act synergistically in driving reflection 
asymmetry, necessitating a revised paradigm for understanding octupole 
correlation. 
\end{abstract}

\maketitle

\section{Introduction}

Octupole collectivity is among the most intriguing symmetry-breaking modes in 
atomic nuclei and has long been an important topic in nuclear-structure 
physics~\cite{Nazarewicz94,Butler96,Butler16,Butler20,Wang22}. It is also 
important in searches for physics beyond the Standard Model~\cite{Engel13}. 
As a finite many-body quantum system, the nucleus develops its ground-state 
shape through the interplay between shell effects and residual interactions, 
and modern self-consistent energy-density-functional frameworks provide a 
unified microscopic description of such emergent deformation 
phenomena~\cite{Bohr75,Frauendorf01,ring80,Bender03}. Early theoretical 
studies already indicated that stable reflection-asymmetric shapes and 
parity-doublet structures can occur in specific regions of the nuclear chart, 
notably in the actinide region~\cite{Nazarewicz84,Leander86}. Extensive 
spectroscopic investigations have since established the systematics of 
octupole collectivity across several mass regions, with representative 
examples in even-even nuclei~\cite{Huang16,Zhu2020,Bucher16, 
Bucher17,Urban91, Iacob96,Chen06,Gaffney13, Cocks97, ButlerRa20}, odd-odd 
nuclei~\cite{Sheline88,Sheline89,Liu16,Xu22,Xu24,Xiao22} and odd-$A$ 
nuclei~\cite{Fernandez91,Sheline86,Reich86,Dahlinger88,Morse20,Zhu01,Zhu99,UrbanLa96}.

\begin{figure}[H]
    \centering
    \includegraphics[height=16cm]{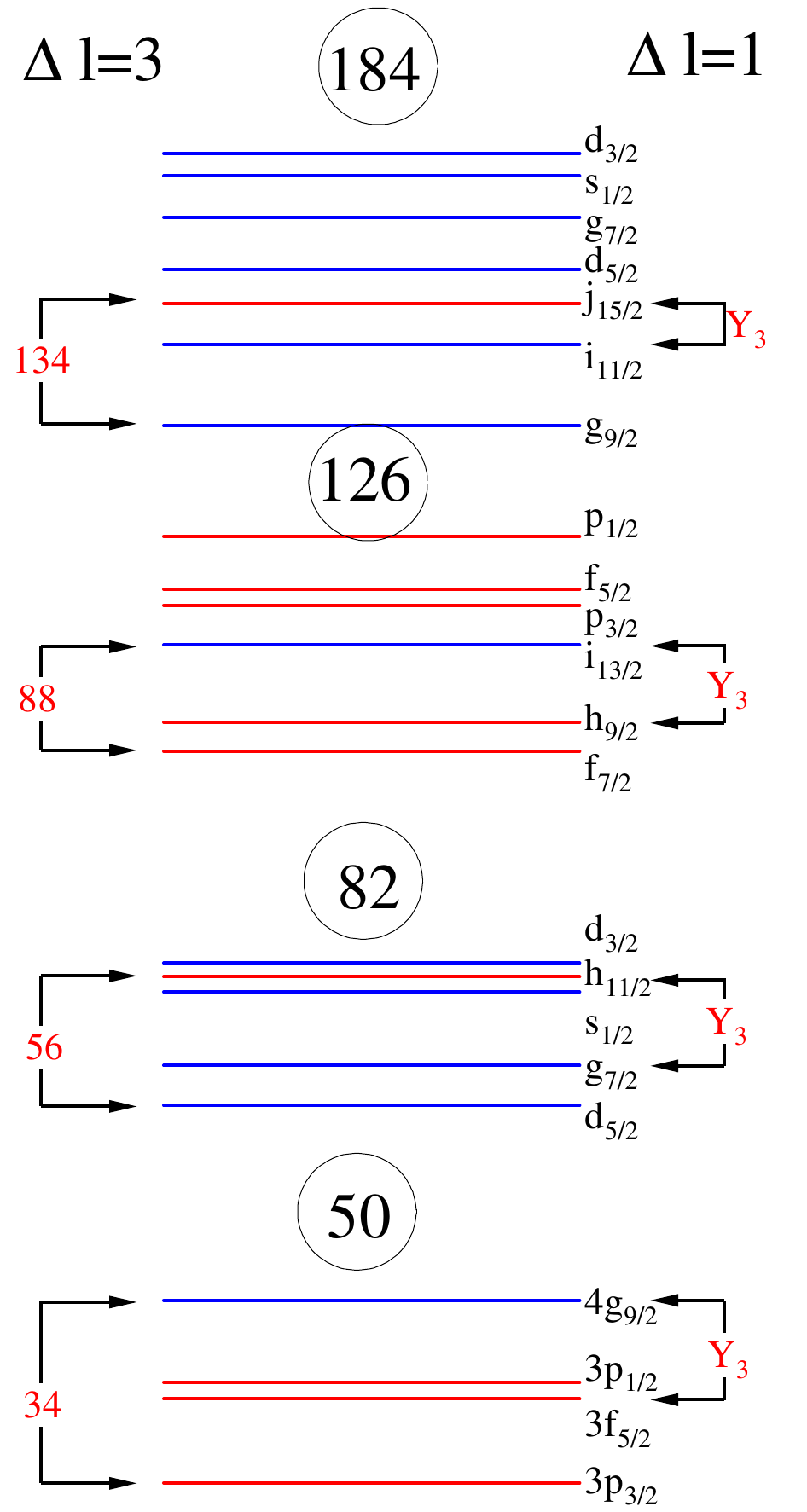}
    \caption{Spherical neutron single-particle levels and the dominant octupole-driving couplings. Blue (red) lines denote positive- (negative-) parity
    orbitals. The arrows on the left and right sides of the levels indicate
    octupole couplings between opposite-parity orbitals with $\Delta l=3$   and
    $\Delta l=1$, respectively. For a similar schematic focusing on the
    $\Delta l=3$ driven picture, see Refs.~\cite{Butler96,Butler16}.}
    \label{fig:coupling}
\end{figure}

Microscopically, octupole correlations are favored when opposite-parity 
single-particle orbitals with large octupole matrix elements lie near the 
Fermi surface. From a spherical-shell perspective, the strongest driving 
couplings have often been associated with $\Delta l=\Delta j=3$, where $j$ 
and $l$ denote the total and orbital angular momenta of the particles, as 
illustrated in the left coupling of Fig.~\ref{fig:coupling}. This naturally 
leads to the well-known octupole-favorable nucleon numbers around $\sim 
34\,(g_{9/2}\leftrightarrow p_{3/2})$, $\sim 56\,(h_{11/2}\leftrightarrow 
d_{5/2})$, $\sim 88\,(i_{13/2}\leftrightarrow f_{7/2})$, and $\sim 
134\,(j_{15/2}\leftrightarrow 
g_{9/2})$~\cite{Butler96,Butler16,Butler20,Nazarewicz94,Huang16,Chen06,Cocks97,ButlerRa20,Iacob96,Gaffney13,Agbemava16,Guzman21,Han23}.

%Quadrupole deformation reshapes the single-particle spectrum and pairing 
%correlations smear occupation probabilities around the Fermi energy. As a 
%result, additional opposite-parity orbitals can become quasi-degenerate and 
%contribute significantly to octupole-induced 
%mixing~\cite{Agbemava16,Guzman21}. 
%Correspondingly, octupole phenomena are systematically observed in these mass 
%regions. 

Octupole correlations mix orbitals via the operator $r^2 Y_{3\nu}$, with the 
corresponding matrix element $\bra{N'l'j'\Omega'}r^2Y_{3\nu}\ket{Nlj\Omega}$, 
where $\ket{Nlj\Omega}$ denotes the eigenstate of the spherical 
single-particle Hamiltonian. The parity selection rule requires $\Delta l$ to 
be odd, meaning that both $\Delta l=1$ and $\Delta l=3$ couplings are 
allowed~\cite{Nilsson55,sakurai}. Consequently, a microscopic understanding 
of octupole collectivity should not rely solely on the traditional $\Delta 
l=\Delta j=3$ picture, but also on the often overlooked $\Delta l=1$ 
component.

Fig.~\ref{fig:coupling} provides a schematic illustration of spherical 
neutron single-particle levels and the corresponding octupole coupling 
patterns. Blue (red) lines denote positive (negative) parity orbitals. The 
arrows on the left and right sides of the levels indicate octupole couplings 
between opposite-parity orbitals with $\Delta l=3$ and $\Delta l=1$, 
respectively. It should be noted that the $\Delta l=1$ partners have smaller 
energy splittings, which might facilitate octupole-induced mixing. 

The present work aims to clarify, in a quantitative manner, how the $\Delta 
l=1$ mode contributes to octupole-driven parity mixing in realistic nuclei. 
To address this issue, we take orbitals near the octupole magic number $N = 
134$ as a benchmark, and  systematically evaluate the octupole coupling 
contributions.  The impact of $\Delta l=1$ and $\Delta l=3$ couplings on the 
intrinsic single-particle structure and the rotational structure is analyzed 
within a reflection-asymmetric Nilsson potential and particle-rotor model. 

%These nuclei are chosen as a representative benchmark rather than a unique 
%special case, the same physical competition and cooperation between $\Delta 
%l=1$ and $\Delta l=3$ octupole couplings is expected to persist in other 
%octupole-magic regions. 

\section{Theoretical framework}

The reflection-asymmetric Nilsson~\cite{Nilsson55,Nilsson69,Wang19}  and 
particle-rotor models~\cite{Leander84,Leander87,Leander88,Wang19} are used in 
this work, which could provide a transparent framework to disentangle 
single-particle components and rotational structure, enabling a quantitative 
assessment of the $\Delta l=1$ coupling and its impact on rotational 
properties.

\subsection{Reflection-asymmetric potential}

The intrinsic Hamiltonian $\hat{H}_{\text{intr}}$ for the valence nucleon is
\begin{equation}
\hat{H}_{\text{intr}}=\hat{H}_{\text{sp}}+\hat{H}_{\text{pair}}=  \sum_{\nu>0}\left(e_\nu-\lambda\right)\left(a_\nu^{\dagger} a_\nu+a_{\bar{\nu}}^{\dagger} a_{\bar{\nu}}\right) -\frac{\Delta}{2} \sum_{\nu>0}\left(a_\nu^{\dagger} a_{\bar{\nu}}^{\dagger}+a_{\bar{\nu}} a_\nu\right)
\end{equation}
where $\lambda$ denotes the Fermi energy, $\Delta$ is the pairing-gap 
parameter,  and $\ket{\bar{\nu}}$ denotes the degenerate partner state 
associated with $\ket{\nu}$. The operators $a_\nu^{\dagger}$ ($a_\nu$) create 
(annihilate) a nucleon in the single-particle state $\ket{\nu}$.  The 
single-particle energy $e_\nu$ is obtained by diagonalizing the Hamiltonian 
$\hat{H}_{\text{sp}}$. In this work, we employ the Nilsson-type 
single-particle Hamiltonian \cite{Nilsson55,Nilsson69}, 
\begin{equation}
            \hat{H}_{\text{sp}}=h_0+V(r, \theta, \varphi)=-\frac{1}{2} \hbar \omega_0\nabla^2-\kappa \hbar \omega_0\left\{2\vec{l} \cdot \vec{s}+\mu\left(\vec{l}^2-\braket{\vec{l}^2}_N\right)\right\}+V(r, \theta, \varphi),    \label{eq:nilsson}
\end{equation}
where $\omega_0$ is the oscillator frequency, 
$-\frac{1}{2}\hbar\omega_0\nabla^2$ is the kinetic-energy term, 
$\vec{l}\cdot\vec{s}$ is the spin--orbit term, and 
$\left[\vec{l}^2-\braket{\vec{l}^2}_N\right]$ is an empirical 
$l^2$-correction that introduces $l$-dependent level shifts.  The parameters 
$\kappa$ and $\mu$ are taken from Ref.~\cite{Bengtsson85}.

The reflection-asymmetric axially deformed potential $V(r,\theta,\varphi)$ is 
written as~\cite{Wang19,Hamamoto91} 
\begin{equation}
        V(r, \theta, \varphi)=\hbar \omega_{0} r^{2}\left[\frac{1}{2}-\beta_{2} Y_{20}-\beta_{3} Y_{30}\right].
\end{equation}
Here $\beta_2$ and $\beta_3$ are the axial quadrupole and octupole deformation parameters, respectively.

The single-particle states can be expanded in the eigenbasis of $h_0$, i.e., the spherical harmonic-oscillator basis $\ket{Nlj\Omega}$ \cite{Nilsson55},
\begin{equation}
    \ket{Nlj\Omega}=\ket{N l} \ket{lj\Omega}=R_{n_r l} \ket{lj\Omega}
\end{equation}
where $n_r$ is the radial quantum number and $R_{n_r l}(r)$ is the radial wave function. The spin--angular part reads
\begin{equation}
    \begin{aligned}
        \ket{lj\Omega}= \sum\limits_{m,m_s}\braket{lmsm_s\mid j\Omega}Y_{lm}\phi_{s,m_s}
        =           
             t\sqrt{\frac{l+t\Omega+\frac{1}{2}}{2l+1}}Y_{l\Omega-\frac{1}{2}}\phi_{\frac{1}{2},\frac{1}{2}}+\sqrt{\frac{l-t\Omega+\frac{1}{2}}{2l+1}}Y_{l\Omega+\frac{1}{2}}\phi_{\frac{1}{2},-\frac{1}{2}}         
    \end{aligned}
\end{equation}
where   
\begin{equation}
    t=2(j-l)=\left\{\begin{aligned}
        & 1,      &  & j=l+\frac{1}{2}, \\
        & -1, &  & j=l-\frac{1}{2}.
    \end{aligned}\right.
\end{equation}
Here $Y_{lm}$ is the spherical harmonic describing the angular motion,  and 
$\phi_{s,m_s}$ is the spin wave function (for a nucleon, $s=\tfrac{1}{2}$). 
The principal quantum number is $N=2n_r+l$, the total angular momentum 
quantum number is $j$, and $\Omega$ denotes the projection of $\vec{j}$ onto 
the quantization symmetry axis. 

In the deformed potential $V(r,\theta,\varphi)$, the matrix element of the $r^2Y_{\mu\nu}$ term factorizes as
\begin{equation}
\bra{N'l'j'\Omega'}r^2Y_{\mu\nu}\ket{Nlj\Omega}=\bra{N'l'}r^2\ket{Nl}\,\bra{l'j'\Omega'}Y_{\mu\nu}\ket{lj\Omega},
\end{equation}
where $\bra{N'l'}r^2\ket{Nl}$ is the radial part and $\bra{l'j'\Omega'}Y_{\mu\nu}\ket{lj\Omega}$ is the angular part.
Using the integral formula for the product of three spherical harmonics \cite{Nilsson55}, the angular matrix elements relevant to the axial octupole and quadrupole deformations are
\begin{equation}
    \begin{aligned}
                     \braket{l'j'\Omega'|Y_{30}|lj\Omega}           &= \sqrt{\frac{7}{4\pi}}\frac{\braket{ 30l0\mid l'0}}{2l'+1}\\
        &\times  (    \sqrt{(l'-t'\Omega'+\frac{1}{2})(l-t\Omega+\frac{1}{2})}\braket{ 3 0 l \Omega+\frac{1}{2} \mid l' \Omega'+\frac{1}{2}}\\
        &+       t't  \sqrt{(l'+t'\Omega'+\frac{1}{2})(l+t\Omega+\frac{1}{2})}\braket{ 3 0 l \Omega-\frac{1}{2} \mid l' \Omega'-\frac{1}{2}})
    \end{aligned}
\end{equation}
\begin{equation}
    \begin{aligned}
                     \braket{l'j'\Omega'|Y_{20}|lj\Omega}           &= \sqrt{\frac{5}{4\pi}}\frac{\braket{ 20l0\mid l'0}}{2l'+1}\\
        &\times  (    \sqrt{(l'-t'\Omega'+\frac{1}{2})(l-t\Omega+\frac{1}{2})}\braket{ 2 0 l \Omega+\frac{1}{2} \mid l' \Omega'+\frac{1}{2}}\\
        &+       t't  \sqrt{(l'+t'\Omega'+\frac{1}{2})(l+t\Omega+\frac{1}{2})}\braket{ 2 0 l \Omega-\frac{1}{2} \mid l' \Omega'-\frac{1}{2}})
    \end{aligned}
\end{equation}
where $\braket{\cdots\mid\cdots}$ denote the Clebsch--Gordan coefficients. 
The  expression of the radial matrix elements  $\bra{N'l'}r^2\ket{Nl}$ can be 
found in Ref.~\cite{Nilsson55}.

The intrinsic single-particle eigenstate can be expanded in the spherical
harmonic-oscillator basis as
\begin{equation}
    a_\nu^{\dagger}\ket{0}=\sum_{Nlj} c_{Nlj\Omega}^{(\nu)}\,\ket{Nlj\Omega}.
\end{equation}
 For an axially symmetric
field, $\Omega$ is a good quantum number and takes half-integer values. We 
denote by $a_{\bar{\nu}}^{\dagger}\ket{0}$ the degenerate partner state with 
the opposite projection $-\Omega$ (i.e., $e_{\bar{\nu}}=e_{\nu}$).

To include pairing effects in the model, the single-particle state
$a_\nu^{\dagger}\ket{0}$ should be replaced by the BCS quasiparticle state
$\alpha_\nu^{\dagger}\ket{\tilde{0}}$, where $\ket{\tilde{0}}$ is the BCS vacuum
state. The quasiparticle operators $\alpha_\nu^{\dagger}$ are defined by
\begin{equation}
    \binom{\alpha_\nu^{\dagger}}{\alpha_{\bar{\nu}}}=\left(\begin{array}{cc}
u_\nu & -v_\nu \\
v_\nu & u_\nu
\end{array}\right)\binom{a_\nu^{\dagger}}{a_{\bar{\nu}}},
\end{equation}
where $v_\nu$ is the occupation factor of the state $\nu$, and
$u_\nu^2+v_\nu^2=$1. Meanwhile, the single-particle energies $e_\nu$ should be
replaced by quasiparticle energies
$e_\nu^{\prime}=\sqrt{\left(e_\nu-\lambda\right)^2+\Delta^2}$.

\subsection{Particle-rotor model with octupole deformation}
The total Hamiltonian of the particle-rotor model with octupole deformation 
(PRM) can be expressed as~\cite{Zhang22,Wang19} 
\begin{equation}
    \hat{H}=\hat{H}_{\text {coll }}+\hat{H}_{\text {intr }},
\end{equation}
For an axially deformed system, the collective Hamiltonian takes the form
\begin{equation}
    \hat{H}_{\text {coll }}=\frac{\hat{I}^2-\hat{I}_3^2}{2 \mathcal{J}}+\frac{\hat{j}^2-\hat{j}_3^2}{2 \mathcal{J}}-\frac{\hat{I}_+ \hat{j}_-+\hat{I}_- \hat{j}_+}{2 \mathcal{J}}+\frac{1}{2} E\left(0^{-}\right)\left(1-\hat{P}\right).
\end{equation}
Here $\hat{I}$ and $\hat{j}$ denote the angular momenta of the nucleus and the
valence nucleon, respectively, and $\hat{I}_3$ and $\hat{j}_3$ are their
projections on the symmetry axis.
In the last term, the core parity-splitting parameter $E\left(0^{-}\right)$ 
can be interpreted as the excitation energy of the virtual $0^{-}$ 
state~\cite{Wang19}. The core parity operator is defined as 
$\hat{P}=\hat{\pi}\hat{p}$, where $\hat{\pi}$ and $\hat{p}$ are the 
single-particle and total parity operators, respectively. $\mathcal{J}$ is 
the moment of inertia for rotation about an axis perpendicular to the 
symmetry axis.

To obtain the 
PRM solutions, the total Hamiltonian must be diagonalized in a complete basis 
space, which couples the rotation of the core with the intrinsic wave 
function of a quasiparticle. For octupole-deformed nuclei, parity symmetry is 
broken in the intrinsic frame but restored in the laboratory frame. 
Therefore, the intrinsic wave function must be constructed with definite 
parity via parity projection 
\begin{equation}
    \begin{aligned}
        \psi_{+}^\nu & =(1+\hat{p}) \alpha_\nu^{\dagger}\ket{\tilde{0}} \Phi_a=\left(1+\hat{P} \hat{\pi}\right) \alpha_\nu^{\dagger}\ket{\tilde{0}}  \Phi_a \\
        \psi_{-}^\nu & =(1-\hat{p}) \alpha_\nu^{\dagger}\ket{\tilde{0}} \Phi_a=\left(1-\hat{P} \hat{\pi}\right) \alpha_\nu^{\dagger}\ket{\tilde{0}}  \Phi_a \\
    \end{aligned}
    \label{eq:intrinsic}
\end{equation}
Here $\alpha_\nu^{\dagger}\ket{\tilde{0}} \Phi_a$ is the strong-coupled 
intrinsic core-quasiparticle wave function; $\Phi_a$ represents that the core 
has the same orientation in space as the intrinsic single-particle potential, 
and $\alpha_\nu^{\dagger}\ket{\tilde{0}}$ is the BCS quasiparticle state of 
the valence nucleon. 
The final symmetrized strong-coupling basis can be written as ~\cite{Wang19} 
\begin{equation}
    \ket{\Psi_{I M K p}^\nu}=\frac{1}{2}\left(1+\hat{R}\right)\ket{I M K} \psi_{ p}^\nu,\qquad p=\pm, 
    \label{eq:basis}
\end{equation}
where $\hat{R}$ means rotation of $\pi$ around an axis perpendicular to the 
symmetry axis. The system also satisfies the symmetry 
$\hat{S}=\hat{p}\hat{R}$, which is the reflection operator with respect to a 
principal plane containing the symmetry axis. The intrinsic states are then 
projected onto good parity to restore the symmetry in the laboratory frame, 
allowing a consistent description of parity doublets and electromagnetic 
transitions.  By diagonalization, the wave functions of the Hamiltonian can 
be written as 
\begin{equation}
    \ket{I M p}=\sum_{K \nu} C_{K \nu}^{I p} \ket{\Psi_{I M K p}^\nu},\qquad p=\pm.
\end{equation}

In this work, the dipole and quadrupole electric transitions ($E1$ and $E2$) 
in the axially symmetric case are important. The decay rate is written 
as~\cite{Bohr75} 
\begin{equation}
T(E\lambda,I_i\to I_f)=\frac{8 \pi(\lambda+1)}{\lambda[(2 \lambda+1)!!]^{2}} \frac{1}{\hbar}\left(\frac{\omega}{c}\right)^{2 \lambda+1} B(E\lambda,I_i\to I_f),\quad \lambda=1,2,
\label{Eq:T}
\end{equation}
with the reduced transition probability
\begin{equation}
B(E\lambda,I_i\to I_f)=\frac{1}{2I_i+1}\left|\braket{ I_f\|\mathcal{M}_{\lambda}\|I_i}\right|^2,
\label{Eq:B-RM}
\end{equation}
in terms of the matrix elements of the electric multipole operator
$\mathcal{M}_{\lambda \mu}$ of order $\lambda,\mu$ between an initial
state $\ket{I_iM_ip_i}$ and a final state $\ket{I_fM_fp_f}$. Using the
Wigner-Eckart theorem, one obtains
\begin{equation}
B(E\lambda, I\to I')=\frac{2 I^{\prime}+1}{2 I+1}\cdot \frac{\bra{I'M'p'}\mathcal{M}_{\lambda \mu}\ket{IMp}^{2}}{\braket{IM\lambda\mu|I'M'}^2}.
\label{Eq:B}
\end{equation}

For axially deformed nuclei, the multipole operators in the laboratory frame ($\mathcal{M}_{\lambda
\mu}$) and the intrinsic system ($\mathcal{M}_{\lambda
\mu}^{'}$) are connected by the relation~\cite{Bohr75},
\begin{equation}
\mathcal{M}_{\lambda \mu}=D^{\lambda}_{\mu 0}\mathcal{M}_{\lambda 0}^{'}=D^{\lambda}_{\mu 0}Q_{\lambda0}^{'},\quad \lambda=1,2.
\end{equation}
According to the empirical equations, the intrinsic quadrupole moment is 
written as $ Q_{20}^{'}=\frac{3e}{4\pi}ZR_0^{2}\beta_2$~\cite{Bohr75}, and 
the electric dipole moment is $Q_1=c_1AZ\text{e}\beta_2\beta_3$ with 
$c_1=0.00069~\text{fm}$~\cite{Strutinsky66,Leander86}.

\subsection{Numerical details}

Based on the reflection-asymmetric Nilsson model, the $\Delta l = 1$ and 
$\Delta l = 3$ mixing ratios are systematically evaluated, taking orbitals 
around the octupole magic number $N = 134$ as a benchmark. The adopted 
deformation ranges are $\beta_2=0$ and $0.15$, and $\beta_3=0\sim0.15$. 

Based on the PRM, we take $^{221}$Ra and $^{223}$Th as 
examples to demonstrate the impact of the $\Delta l = 1$ coupling on rotational 
structure. For $^{221}$Ra, covariant density functional theory calculations 
on a three-dimensional lattice indicate a stable reflection-asymmetric shape 
with $(\beta_2,\beta_3)=(0.15,0.11)$~\cite{zhang23}.
Thus, we adopt $\beta_2=0.15$ and $\beta_3=0.10$ in the present PRM calculations 
for the $N=133$ isotones $^{221}$Ra and $^{223}$Th. The neutron Fermi energy 
is set to $\lambda=49.84\mathrm{MeV}$, which corresponds to the 
67th single-particle level in both $^{221}$Ra and $^{223}$Th. 
The single-particle basis includes five levels above and five levels 
below the Fermi level. The pairing correlation is taken into account by the 
empirical formula $\Delta=12/ \sqrt{A} \mathrm{MeV}$.

For the core, a spin-dependent moment of inertia (MoI),
$\mathcal{J}(I)=(a+b I) \hbar^2/ \mathrm{MeV}$, is required to reproduce the
experimental energy spectra.
$(a,b)=(30,2.8)$ is adopted for both the positive- and negative-parity bands 
for $^{221}$Ra, and $(a,b)=(35,2.35)$ for $^{223}$Th. The core parity 
splitting parameter $E\left(0^{-}\right)=0.15$ MeV for $^{221}$Ra and $0.207$ 
MeV for $^{223}$Th is obtained by taking the average experimental $1^{-}$ 
excitation energy in the two neighboring even-even nuclei~\cite{Leander88}. A 
Coriolis attenuation factor $\xi$ is introduced in the PRM description, with 
$\xi=0.50$ (positive parity) and $0.65$ (negative parity) for $^{221}$Ra, and 
$\xi=0.50$ (positive parity) and $0.75$ (negative parity) for $^{223}$Th.

For electric transitions, the intrinsic dipole moment is taken as 
$Q_{10}=c_1AZ\text{e}\beta_2\beta_3=0.2\,e\mathrm{fm}$ for both $^{221}$Ra 
and $^{223}$Th with $c_1=0.00069~\text{fm}$. The intrinsic quadrupole moment 
is set to $Q_{20}=\frac{3e}{4\pi}ZR_0^{2}\beta_2=170\,e\mathrm{fm}^2$, with 
$R_0=1.2A^{1/3}\,\mathrm{fm}$~\cite{Wang19}. 

\section{Discussion}

\subsection{Single-Particle Levels}

\begin{figure}[H]
    \centering
    \includegraphics[height=8cm]{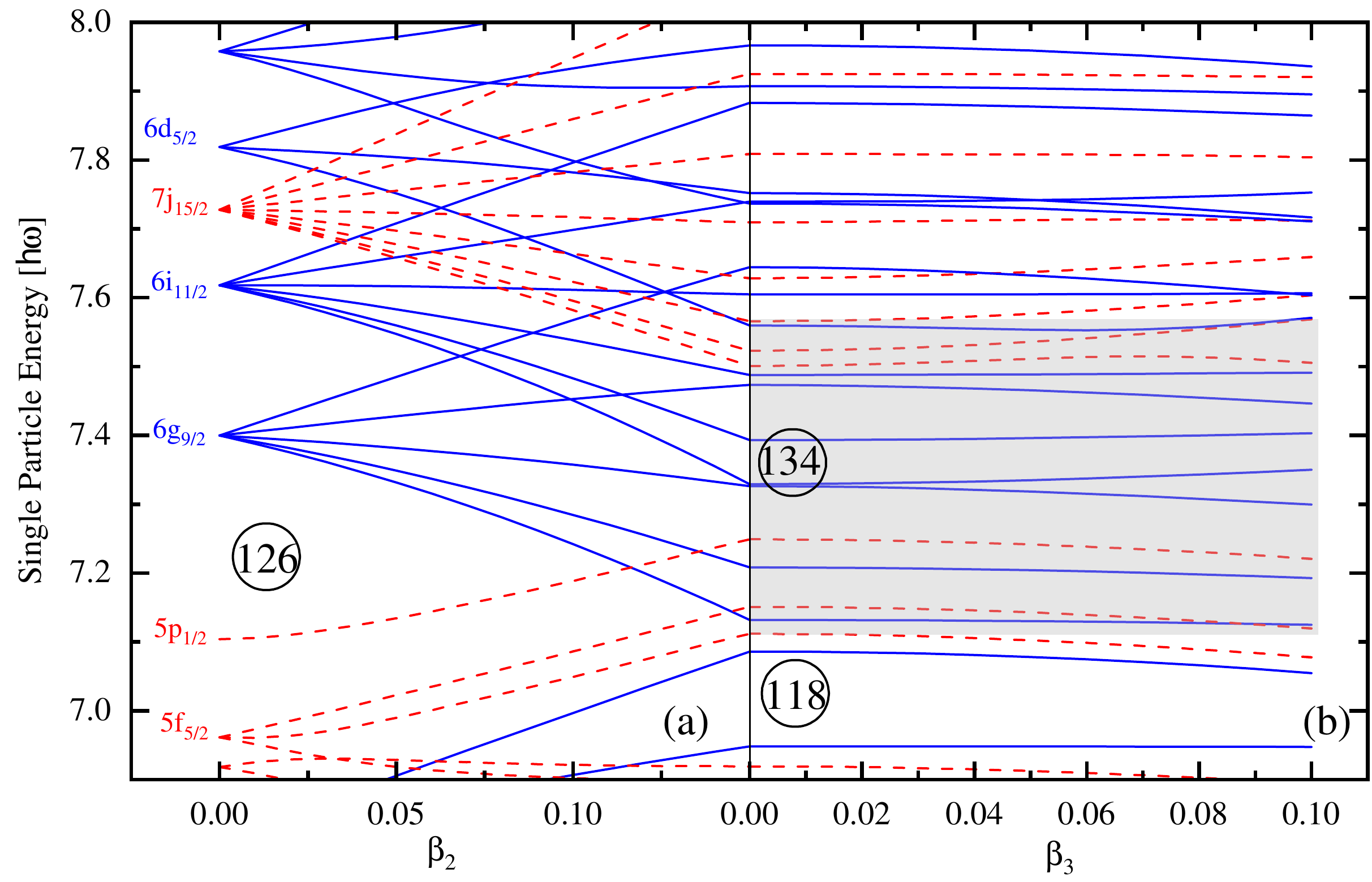}
    \caption{The neutron single-particle levels, obtained by diagonalizing the
    reflection-asymmetric axial Nilsson Hamiltonian as functions of the $\beta_2$ ($\beta_3=0$) 
    and $\beta_3$ ($\beta_2=0.15$).  In panel (a), each level has a definite parity, and the blue solid lines represent
    positive-parity states, while the red dashed lines represent
    negative-parity states. In
    panel (b), octupole deformation mixes states of opposite parity, the colors  are used to indicate the parity of the dominant components. The shaded region marks the neutron single-particle levels
    included in the PRM model space. }
    \label{fig:nilsson}
\end{figure}

The neutron single-particle levels are obtained by diagonalizing the
reflection-asymmetric axial Nilsson Hamiltonian in Eq.~(\ref{eq:nilsson}).
Fig.~\ref{fig:nilsson} displays these levels as functions of the deformation
parameters. The quasiparticle states are subsequently derived from these
single-particle states via the BCS approximation at the deformation
$\beta_2=0.15, \beta_3=0.1$, corresponding to the deformation of $^{221}$Ra and $^{223}$Th.

In Fig.~\ref{fig:nilsson}, the neutron single-particle levels are plotted as 
functions of (a) quadrupole deformation $\beta_2$ (with $\beta_3=0$) and (b) 
octupole deformation $\beta_3$ (with $\beta_2=0.15$). To ensure continuity of
the level trajectories, the states are traced diabatically as functions of
deformation. The shaded region  
marks the neutron single-particle levels included in the PRM model space. The 
blue solid lines represent positive-parity states, while the red dashed lines 
represent negative-parity states. In panel (a), each level has a definite 
parity. In panel (b), octupole deformation mixes states of opposite parity, 
such that each level contains both parity components. Due to the relatively 
small octupole deformation considered here, the admixture of the 
opposite-parity component remains weak; therefore, the same color scheme as 
in panel (a) is used to indicate the dominant parity component.  The octupole 
coupling near $N=134$ is generally attributed to the mixing between the 
$j_{15/2}$ and $g_{9/2}$ orbitals. While the single-particle level diagram in 
Fig.~\ref{fig:nilsson} reveals the energy ordering, it does not provide 
insight into the detailed wave function composition or the direct 
identification of contributions from different coupling modes. 

\subsection{Wave function Mixing}

The single-particle level diagrams in Fig.~\ref{fig:nilsson} illustrate the 
energy ordering under octupole deformation but do not reveal the detailed 
wave function composition. To gain a deeper insight into the microscopic 
origin of octupole-induced parity mixing, we focus on the typical 
octupole-deformed nuclei $^{221}$Ra and $^{223}$Th ($\beta_2=0.15, 
\beta_3=0.1$) and analyze in detail the wave function components of the 
single-particle orbitals near the Fermi surface. 
Table~\ref{tab:wavefunctions} lists the main components with coefficient 
absolute values greater than 0.1 for the 11 pairs of degenerate levels (level 
indices $k=62\text{--}72$) corresponding to neutron numbers from $N = 123$ to 
$144$. 

\begin{table}[H]
    \centering
    \small
    \caption{The main components of single-particle wave functions for
    orbitals within five levels above and below the Fermi surface of
    $^{221}$Ra and  $^{223}$Th at $\beta_2=0.15, \beta_3=0.1$.}
    \label{tab:wavefunctions}
    \begin{tabular}{c|c|c|l}
        \hline
        Level index $k$& Neutron number & $\Omega$ & Wave function\\  
        \hline
        62 & 123\&124 & 5/2      & $0.93\ket{ 5f_{ 5/2} } -0.21\ket{ 6g_{ 9/2} } -0.17\ket{ 5h_{ 9/2} } +0.16\ket{ 6i_{11/2} }$\\ \hline
        63 & 125\&126 & 1/2      & $\begin{aligned}  &0.65\ket{ 6g_{ 9/2} } +0.37\ket{ 6d_{ 5/2} } -0.33\ket{ 5p_{ 3/2} } -0.32\ket{ 6i_{13/2} } +0.25\ket{ 5p_{ 1/2} }\\& +0.23\ket{ 7j_{15/2} } +0.20\ket{ 7h_{11/2} } +0.13\ket{ 6s_{ 1/2} } +0.11\ket{ 8i_{13/2} }\end{aligned}$\\ \hline
        64 & 127\&128 & -3/2     & $\begin{aligned}  &0.68\ket{ 6g_{ 9/2} } -0.39\ket{ 5p_{ 3/2} } +0.36\ket{ 6d_{ 5/2} } -0.29\ket{ 6i_{13/2} }\\& +0.26\ket{ 7j_{15/2} } +0.17\ket{ 7h_{11/2} } +0.11\ket{ 8i_{13/2} }\end{aligned}$\\ \hline
        65 & 129\&130 & 1/2      & $\begin{aligned}  &0.78\ket{ 5p_{ 1/2} } -0.40\ket{ 5f_{ 5/2} } +0.37\ket{ 5p_{ 3/2} } -0.21\ket{ 6d_{ 5/2} }\\& +0.12\ket{ 6g_{ 7/2} } -0.11\ket{ 6i_{11/2} }\end{aligned}$\\ \hline
        66 & 131\&132 & 5/2      & $\begin{aligned}  &0.81\ket{ 6g_{ 9/2} } +0.32\ket{ 7j_{15/2} } -0.25\ket{ 6i_{13/2} } +0.22\ket{ 5f_{ 5/2} }\\& +0.20\ket{ 6d_{ 5/2} } -0.17\ket{ 6g_{ 7/2} } -0.16\ket{ 6i_{11/2} } +0.10\ket{ 8i_{13/2} }\end{aligned}$\\ \hline
        67 & 133\&134 & 1/2      & $\begin{aligned}  &0.77\ket{ 6i_{11/2} } +0.44\ket{ 6g_{ 7/2} } -0.27\ket{ 5f_{ 5/2} } +0.19\ket{ 6d_{ 3/2} }+0.13\ket{ 7j_{13/2} }\\&  +0.12\ket{ 8k_{15/2} } -0.12\ket{ 5h_{ 9/2} } +0.11\ket{ 6g_{ 9/2} } -0.10\ket{ 4g_{ 7/2} }\end{aligned}$\\ \hline
        68 & 135\&136 & -3/2     & $\begin{aligned}  &0.85\ket{ 6i_{11/2} } +0.38\ket{ 6g_{ 7/2} } -0.18\ket{ 5f_{ 5/2} } -0.18\ket{ 6g_{ 9/2} }\\& +0.12\ket{ 8k_{15/2} } -0.10\ket{ 4g_{ 7/2} }\end{aligned}$\\ \hline
        69 & 137\&138 & -7/2     & $0.84\ket{ 6g_{ 9/2} } +0.38\ket{ 7j_{15/2} } +0.28\ket{ 6i_{11/2} } -0.17\ket{ 6i_{13/2} } +0.15\ket{ 5f_{ 7/2} }$\\ \hline
        70 & 139\&140 & 5/2      & $0.91\ket{ 6i_{11/2} } +0.26\ket{ 6g_{ 9/2} } +0.25\ket{ 6g_{ 7/2} } +0.12\ket{ 8k_{15/2} }$\\ \hline
        71 & 141\&142 & 1/2      & $\begin{aligned}  &0.52\ket{ 7j_{15/2} } -0.47\ket{ 6d_{ 5/2} } -0.38\ket{ 6s_{ 1/2} } +0.32\ket{ 6d_{ 3/2} } -0.27\ket{ 6i_{11/2} }\\& +0.23\ket{ 6g_{ 9/2} } -0.21\ket{ 5p_{ 1/2} } -0.13\ket{ 7f_{ 7/2} } -0.12\ket{ 6i_{13/2} }\end{aligned}$\\ \hline
        72 & 143\&144 & -3/2     & $\begin{aligned}  &0.88\ket{ 7j_{15/2} } -0.28\ket{ 6g_{ 9/2} } +0.24\ket{ 7h_{11/2} } -0.20\ket{ 6d_{ 5/2} }\\& -0.12\ket{ 5h_{11/2} } +0.12\ket{ 8k_{17/2} }\end{aligned}$     \\
        \hline
    \end{tabular}   
\end{table}

Here we focus on analyzing the level $k=67$ at the Fermi surface, 
corresponding to nucleon numbers 133 and 134 with $\Omega = 1/2$, which is 
dominated by the following components: 
\begin{equation}
    \ket{\psi_{67}} \approx 0.77\ket{6i_{11/2}} + 0.44\ket{6g_{7/2}} - 0.27\ket{5f_{5/2}} + 0.19\ket{6d_{3/2}} + 0.13\ket{7j_{13/2}} + \cdots
\end{equation}
Among these, $\ket{6i_{11/2}}\leftrightarrow \ket{5f_{5/2}}$,
$\ket{7j_{13/2}}\leftrightarrow \ket{6g_{7/2}}$,
$\ket{5h_{9/2}}\leftrightarrow \ket{6d_{3/2}}$ constitute $\Delta l = 3$
couplings, while $\ket{6g_{7/2}}\leftrightarrow \ket{5f_{5/2}}$,
$\ket{7j_{13/2}}\leftrightarrow \ket{8k_{15/2}}$,
$\ket{6i_{11/2}}\leftrightarrow \ket{5h_{9/2}}$ constitute $\Delta l = 1$
couplings.

To quantitatively characterize parity mixing induced by octupole 
interactions, we define a channel-resolved 
relative mixing ratio 
$M_{\Delta l, \Delta j}^k$ for the $k$-th level as: 
\begin{equation}
    M_{\Delta l, \Delta j}^k = \sum_{\substack{|l-l'|=\Delta l, \\|j-j'|=\Delta j }}\left|  C_{nlj\Omega}^k C_{n'l'j'\Omega'}^k \right|\Bigg/ \sum\left|  C_{nlj\Omega}^k C_{n'l'j'\Omega'}^k \right|,
\label{eq:Mljk}
\end{equation}
where $C_{nlj\Omega}^k$ and $C_{n'l'j'\Omega'}^k$ are the expansion 
coefficients of the $k$-th single-particle wave function on the spherical 
basis states $\ket{n l j \Omega}$ and $\ket{n' l' j' \Omega'}$, respectively. 
This definition directly measures the relative contribution of a given 
$(\Delta l,\Delta j)$ channel to the total wave function mixing, i.e., the 
share of that octupole-allowed channel in the total mixing strength over all 
channels retained in the model space. 
 According to angular momentum coupling 
theory, states of opposite parity can couple through the octupole operator 
$r^2 Y_{30}$, allowing mixing types including: $\Delta l=1, \Delta j=0,1,2$ 
and $\Delta l=3, \Delta j=2,3,4$. We also define the level-averaged mixing ratio 
over $N_{\mathrm{sp}}$ selected single-particle levels at a specific deformation as: 
\begin{equation}
    M_{\Delta l, \Delta j} = \frac{1}{N_{\mathrm{sp}}}\sum_{k=1}^{N_{\mathrm{sp}}} M_{\Delta l, \Delta j}^k,
    \label{eq:Mlj}
\end{equation}
and the corresponding $\Delta j$-summed, level-averaged mixing ratios for $\Delta l=1$ and $\Delta l=3$
as:
\begin{equation}
    \begin{aligned}
        M_{\Delta l=1} &= \sum_{\Delta j=0,1,2} M_{\Delta l=1, \Delta j},\\
        M_{\Delta l=3} &= \sum_{\Delta j=2,3,4} M_{\Delta l=3, \Delta j}.
    \end{aligned}
    \label{eq:Ml}
\end{equation}

\begin{figure}[H]
    \centering
    \includegraphics[height=8cm]{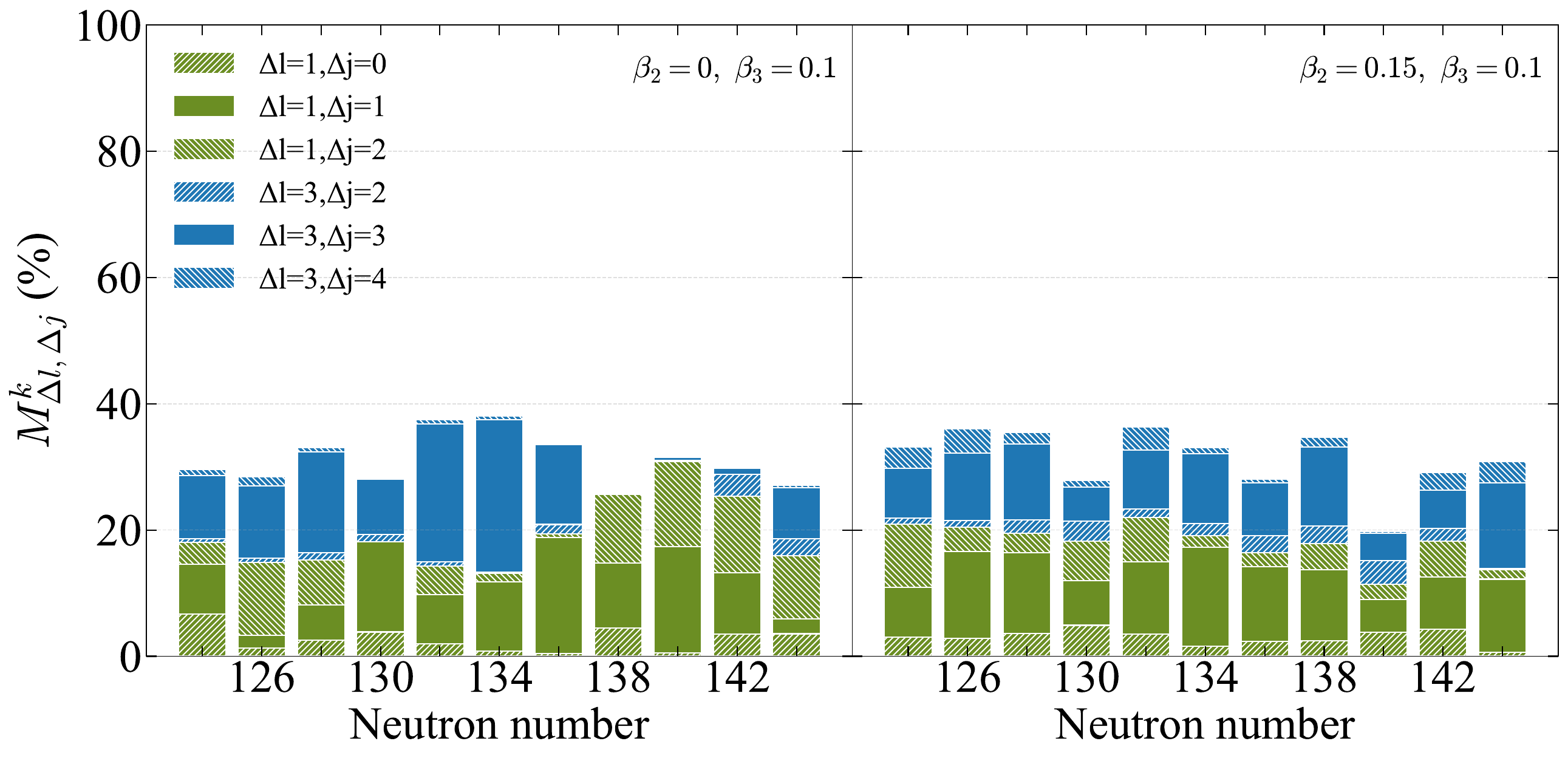}
    \caption{The mixing ratio $M_{\Delta l, \Delta j}^k$ defined in 
    Eq.~(\ref{eq:Mljk}) for neutron single-particle wave functions. The 
    left panel shows the case with pure octupole deformation 
    ($\beta_2=0$, $\beta_3=0.1$). The right panel shows the case with 
    ($\beta_2=0.15$, $\beta_3=0.1$), corresponding to the deformation of 
    $^{221}$Ra and $^{223}$Th. Green (blue) bars denote the $\Delta l=1$ 
    ($\Delta l=3$) components, and different hatching patterns indicate 
    different $\Delta j$ values. } \label{fig:octupole_single} 
\end{figure}

The bar chart in Fig.~\ref{fig:octupole_single} illustrates the mixing 
ratio $M_{\Delta l, \Delta j}^k$ for neutron single-particle wave functions 
corresponding to neutron numbers $123 \sim 144$. For comparison, the left 
panel shows the case with pure octupole deformation ($\beta_2=0$, 
$\beta_3=0.1$), while the right panel shows $\beta_2=0.15$ and $\beta_3=0.1$, 
corresponding to the deformation of $^{221}$Ra and $^{223}$Th.  Green (blue) 
bars denote the $\Delta l=1$ ($\Delta l=3$) components, and different 
hatching patterns indicate different $\Delta j$ values.

As seen in Fig.~\ref{fig:octupole_single}, the level-averaged total octupole-induced mixing 
reaches about 30\% for both deformations $\beta_2=0$ and $\beta_2=0.15$, 
indicating that the overall parity-mixing strength is substantial and 
relatively insensitive to this change in quadrupole shape. At the 
level-by-level scale, two channels, $(\Delta l,\Delta j)=(3,3)$ and $(1,1)$, 
provide the dominant contributions across most orbitals. More importantly, 
the $\Delta l=1$ sector is not a minor correction: its contribution is 
typically comparable to that of $\Delta l=3$, and for $\beta_2=0.15$ it 
becomes larger than the $\Delta l=3$ part for many levels.

A crucial finding is that the $\Delta l=1, \Delta j=1$ coupling provides the 
dominant contribution to the mixing ratio, even surpassing the traditionally
 emphasized $\Delta l=3, \Delta j=3$ coupling. This result directly challenges the conventional view that the 
$\Delta l = 3$ mode provides the dominant contribution to octupole 
collectivity and suggests that $\Delta l=1$ orbital correlations may also 
play a key role in octupole deformation. 

\subsection{Evolution of Different Coupling Components with Deformation}

\begin{figure}[H]
    \centering
    \includegraphics[height=10cm]{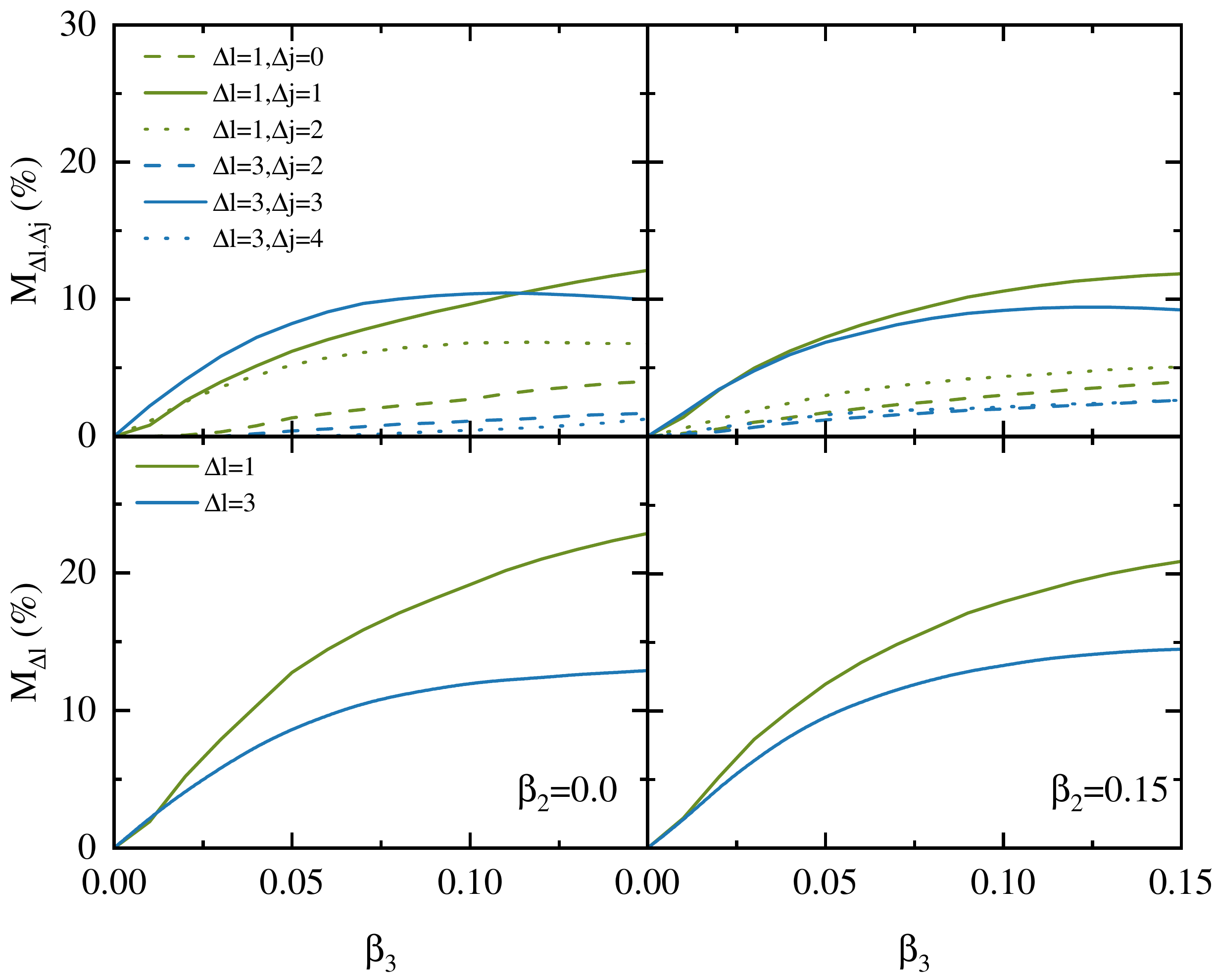}
    \caption{Neutron single-particle wave function mixing ratio as
    functions of $\beta_3$ for $\beta_2=0$ and $0.15$. Upper panels show
    $M_{\Delta l,\Delta j}$, defined in Eq.~(\ref{eq:Mlj}), whereas lower panels show the summed level-averaged ratios
    $M_{\Delta l}$ defined in Eq.~(\ref{eq:Ml}) for the same $\beta_2$ values. Green (blue) curves
    denote the $\Delta l=1$ ($\Delta l=3$) contributions, and different line
    styles indicate different $\Delta j$ values.}
    \label{fig:mixing_ratio}
\end{figure}

To examine the generality of the above conclusion and rule out its dependence 
on specific deformation parameters, we systematically investigate the 
evolution of the mixing ratios with $\beta_3$ for $\beta_2=0$ and 
$0.15$, as shown in Fig.~\ref{fig:mixing_ratio}. The upper panels show the 
component-resolved ratios $M_{\Delta l,\Delta j}$, while the lower panels 
show the corresponding summed level-averaged ratios $M_{\Delta l}$.  The plotted 
quantities are averaged over a fixed near-Fermi set of orbitals defined at 
$\beta_3=0.1$, consisting of five levels below and five levels above the 
Fermi surface. Each selected orbital is then traced diabatically as $\beta_3$ 
varies, and the mixing ratio is evaluated for every tracked level at each 
deformation point.

As shown in the upper panels of Fig.~\ref{fig:mixing_ratio}, for all 
$\beta_2$ values, the mixing ratios show an overall increase with increasing 
$\beta_3$, consistent with the physical picture that larger octupole 
deformation leads to stronger wave function mixing. Moreover, the growth 
rates of the $\Delta l=1$ components (especially $\Delta j=1$) are 
significantly higher than those of the $\Delta l=3$ components. For the case 
of $\beta_2=0$, the $\Delta l=1,\Delta j=1$ contribution exceeds the 
$\Delta l=3,\Delta j=3$ contribution once $\beta_3\gtrsim 0.1$. For 
$\beta_2=0.15$, the $\Delta l=1,\Delta j=1$ contribution remains larger 
than the $\Delta l=3,\Delta j=3$ contribution throughout the entire 
considered $\beta_3$ range. Around 
$\beta_3\approx 0.1$ (close to the deformation of $^{221}$Ra and $^{223}$Th), 
the overall $\Delta l=1$ contribution becomes dominant. The upper panels
also show that the two leading channels are
$(\Delta l,\Delta j)=(3,3)$ and $(1,1)$.

The lower panels of Fig.~\ref{fig:mixing_ratio} provide a direct comparison 
between $M_{\Delta l=1}$ and $M_{\Delta l=3}$. Although both increase with 
$\beta_3$, the $\Delta l=3$ contribution shows a noticeably slower growth 
around $\beta_3\approx 0.1$ for both $\beta_2=0$ and $0.15$, whereas the 
$\Delta l=1$ contribution continues to grow, leading to an increasingly clear 
separation between the two modes.

This systematic behavior indicates that, in realistic nuclei with coexisting
quadrupole and octupole deformations, the $\Delta l=1$ mode is not only
non-negligible but often provides the dominant contribution to parity mixing.

\subsection{Direct Comparison of Octupole Matrix Elements}
\begin{table}[H]
    \centering
    \caption{Matrix elements of the spherical single-particle Hamiltonian $h_0$ within the
    $\Omega=1/2$ subspace. All values are in units of $\hbar\omega_0$.}
    \label{tab:matrix_elements_h0}
    \begin{tabular}{c|cccccccccccc}
        \hline
 & $\ket{5p_{1/2}}$ & $\ket{5p_{3/2}}$ & $\ket{5f_{5/2}}$ & $\ket{6s_{1/2}}$ & $\ket{6d_{3/2}}$ & $\ket{6d_{5/2}}$ & $\ket{6g_{7/2}}$ & $\ket{6g_{9/2}}$ & $\ket{6i_{11/2}}$ & $\ket{6i_{13/2}}$ & $\ket{7h_{11/2}}$ & $\ket{7j_{15/2}}$ \\
\hline
$\braket{h_0}$ & 7.10 & 6.92 & 6.96 & 8.07 & 8.13 & 7.82 & 7.96 & 7.40 & 7.62 & 6.81 & 8.27 & 7.73\\
\hline
    \end{tabular}
\end{table}

\begin{table}[H]
    \centering
    \caption{Matrix elements of the octupole operator $\beta_3 r^2Y_{30}$ within the
    $\Omega=1/2$ subspace. Blank entries denote matrix elements that are
    strictly zero. All values are in units of $\hbar\omega_0$.}
    \label{tab:matrix_elements_b3}
    \begin{tabular}{r|rrrrrrrrrrrr}
\hline
 $\beta_3 r^2Y_{30}$& $\ket{5p_{1/2}}$ & $\ket{5p_{3/2}}$ & $\ket{5f_{5/2}}$ & $\ket{6s_{1/2}}$ & $\ket{6d_{3/2}}$ & $\ket{6d_{5/2}}$ & $\ket{6g_{7/2}}$ & $\ket{6g_{9/2}}$ & $\ket{6i_{11/2}}$ & $\ket{6i_{13/2}}$ & $\ket{7h_{11/2}}$ & $\ket{7j_{15/2}}$ \\
\hline
$\bra{5p_{1/2}}$ &  &  &  &  &  & 0.12 & -0.15 &  &  &  &  &  \\
$\bra{5p_{3/2}}$ &  &  &  &  & 0.12 & -0.07 & 0.03 & -0.15 &  &  &  &  \\
$\bra{5f_{5/2}}$ &  &  &  & 0.08 & -0.05 & 0.03 & -0.08 & 0.02 & -0.13 &  &  &  \\
$\bra{6s_{1/2}}$ &  &  & 0.08 &  &  &  &  &  &  &  &  &  \\
$\bra{6d_{3/2}}$ &  & 0.12 & -0.05 &  &  &  &  &  &  &  &  &  \\
$\bra{6d_{5/2}}$ & 0.12 & -0.07 & 0.03 &  &  &  &  &  &  &  & -0.17 &  \\
$\bra{6g_{7/2}}$ & -0.15 & 0.03 & -0.08 &  &  &  &  &  &  &  & 0.01 &  \\
$\bra{6g_{9/2}}$ &  & -0.15 & 0.02 &  &  &  &  &  &  &  & -0.10 & -0.15 \\
$\bra{6i_{11/2}}$ &  &  & -0.13 &  &  &  &  &  &  &  & 0.01 & 0.01 \\
$\bra{6i_{13/2}}$ &  &  &  &  &  &  &  &  &  &  & -0.06 & -0.11 \\
$\bra{7h_{11/2}}$ &  &  &  &  &  & -0.17 & 0.01 & -0.10 & 0.01 & -0.06 &  &  \\
$\bra{7j_{15/2}}$ &  &  &  &  &  &  &  & -0.15 & 0.01 & -0.11 &  &  \\
\hline
\end{tabular}
\end{table}

To isolate the mechanism of parity mixing, 
Tables~\ref{tab:matrix_elements_h0}--\ref{tab:matrix_elements_b3} list the 
matrix elements of $h_0$ and $H_{\text{oct}}\sim\beta_3 r^2Y_{30}$ at 
$\beta_2=0.15,\beta_3=0.1$. We focus on the $\Omega=1/2$ subspace 
($k=63,65,67,71$ in Table~\ref{tab:wavefunctions}) near Fermi state of 
$N=134$. The sparse nonzero patterns in Tables~\ref{tab:matrix_elements_b3} 
follow the expected angular-momentum and parity selection rules.

As shown in Table~\ref{tab:matrix_elements_h0}, some $\Delta l=1$ pairs can 
even be closer in energy than canonical $\Delta l=3$ partners. For example, 
the traditional $\Delta l=3$ pair $6g_{9/2}\leftrightarrow 7j_{15/2}$ has 
energy difference of $0.33\,\hbar\omega_0$, whereas the $\Delta l=1$ pair 
$6i_{11/2}\leftrightarrow 7j_{15/2}$ has $0.11\,\hbar\omega_0$. As shown in 
Table~\ref{tab:matrix_elements_b3}, at the bare matrix-element level, matrix 
elements for $\Delta l=1$ are comparable in magnitude to those for $\Delta 
l=3$. For example, $\bra{6i_{13/2}} H_{\text{oct}} \ket{7j_{15/2}} = -0.11$, 
while $\bra{6g_{9/2}} H_{\text{oct}} \ket{7j_{15/2}} = -0.15$.  From the 
matrix elements of $h_0$ and $H_{\text{oct}}$, the $\Delta l=1$ coupling  
cannot be regarded as perturbatively negligible.

\subsection{Component-Resolved Single-Particle Octupole Energy Contributions}
To quantify the role of different coupling modes from the perspective of 
energy lowering, we rewrite the single-particle Hamiltonian as 
\begin{equation}
\hat{H}_{\text{sp}}(\beta_3)=\left[h_0+\hbar\omega_0r^2\left(\frac{1}{2}-\beta_2Y_{20}\right)\right]-\beta_3\hat{V}^{(3)},
\qquad
\hat{V}^{(3)}\equiv \hbar\omega_0 r^2Y_{30}.
\end{equation}

The intrinsic single-particle eigenstates satisfy  
$\hat{H}_{\text{sp}}(\beta_3)\ket{\psi_k(\beta_3)}=E_k(\beta_3)\ket{\psi_k(\beta_3)}$, 
which can be expanded in the spherical harmonic-oscillator  
$\ket{\psi_k}=\sum_n C_n^k\ket{n}$ ($\ket{n}$ denotes $\ket{Nlj\Omega}$).

The Hellmann--Feynman theorem gives
\begin{equation}
\frac{dE_k}{d\beta_3}
=
\left\langle \psi_k(\beta_3)\left|\frac{\partial \hat{H}_{\text{sp}}}{\partial \beta_3}\right|\psi_k(\beta_3)\right\rangle
=-\bra{\psi_k(\beta_3)}\hat{V}^{(3)}\ket{\psi_k(\beta_3)}=-\sum_{mn} C_m^{k*}C_n^k\,V_{mn}^{(3)}, 
\label{eq:HF_beta3}
\end{equation}
where $ V_{mn}^{(3)}\equiv \bra{m}\hat{V}^{(3)}\ket{n}$.

To disentangle different angular-momentum coupling modes, we define the 
energy contribution for different $(\Delta l,\Delta j)$ for the $k$-th 
single-particle eigenstate at a fixed deformation $\beta_3$ as 
\begin{equation}
G_{\Delta l,\Delta j}^{k}(\beta_3)
=
\sum_{\substack{l_m,l_n,|l_m-l_n|=\Delta l,\\j_m, j_n|j_m-j_n|=\Delta j}}
\left(C_m^{k*}C_n^k\,V_{mn}^{(3)}\right).
\label{eq:G_dldj}
\end{equation}
The corresponding summed contributions for different single-particle 
eigenstate and different $(\Delta l,\Delta j)$ are defined as 
\begin{equation}
    \begin{aligned}
    G_{\Delta l, \Delta j} &= \sum_k G_{\Delta l, \Delta j}^k,\\
        G_{\Delta l=1} &= \sum_{\Delta j=0,1,2} G_{\Delta l=1, \Delta j},\\
        G_{\Delta l=3} &= \sum_{\Delta j=2,3,4} G_{\Delta l=3, \Delta j}.
    \end{aligned}
    \label{eq:G_dl_sum}
\end{equation}
This definition unifies matrix-element strength and wave function mixing in a 
single energy-based framework, and enables a direct comparison between the 
$\Delta l=1$ and $\Delta l=3$ modes at a given deformation point. In the 
following, we refer to $G_{\Delta l,\Delta j}^{k}$ as the component-resolved 
single-particle octupole energy contribution. It should be emphasized that 
$G_{\Delta l,\Delta j}^{k}$ enters $dE_k/d\beta_3$ with an overall minus 
sign. Therefore, for $\beta_3>0$, a larger positive $G_{\Delta l,\Delta 
j}^{k}$ implies a stronger octupole energy-lowering tendency at the 
single-particle level, whereas a negative value counteracts this tendency. 
When such contributions add coherently over the occupied states, they favor 
the development of stable octupole deformation.

\begin{figure}[H]
    \centering
    \includegraphics[width=12cm]{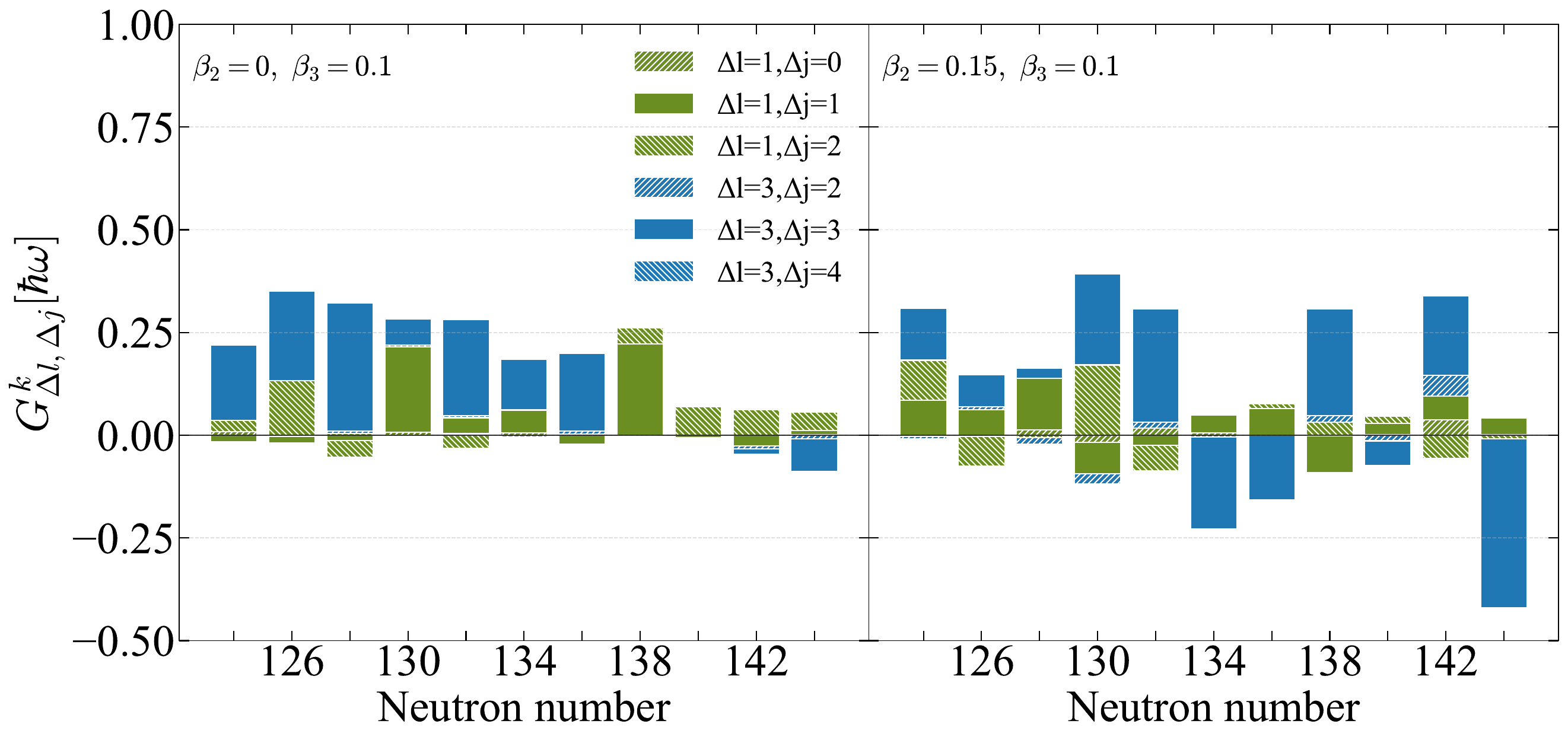}
    \includegraphics[width=12cm]{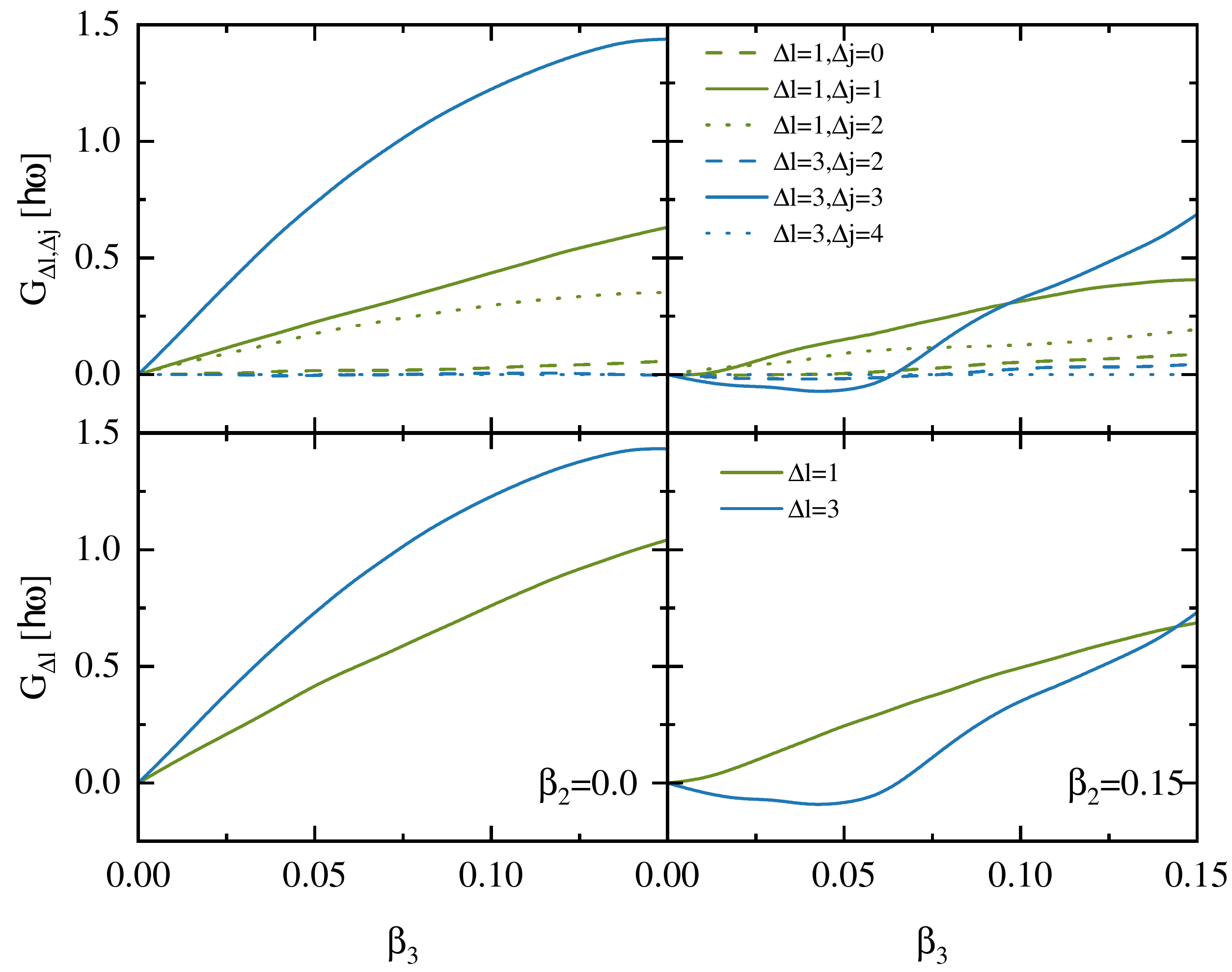}
    \caption{Same plotting scheme as in Fig.~\ref{fig:octupole_single}
    and Fig.~\ref{fig:mixing_ratio}, but with the vertical axis now showing
    the single-particle octupole energy contributions. The upper and lower panels show the
    component-resolved contributions $G^{k}_{\Delta l,\Delta j}$ defined in Eq.~(\ref{eq:G_dldj})
    and the corresponding summed contributions $G_{\Delta l,\Delta j}$ and
    $G_{\Delta l}$ defined in Eq.~(\ref{eq:G_dl_sum}).}
    \label{fig:octupole_single_Energy}
\end{figure}

Fig.~\ref{fig:octupole_single_Energy} shows that the single-particle octupole 
energy contributions are concentrated in two modes, $(\Delta l,\Delta 
j)=(3,3)$ and $(1,1)$, while the remaining modes are subleading.  Such 
behavior is consistent with the mixing ratios shown in  
Fig.~\ref{fig:octupole_single}. From the perspective of energy lowering, the 
$\Delta l=1$ mode is consistently important and, in part of the deformation 
space, even exceeds the $\Delta l=3$ mode. This trend is fully consistent 
with the matrix-element and wave function analyses discussed above. 

The lower panels show that, the net energy-lowering contribution $G_{\Delta 
l}$ from $\Delta l=3$ is larger than that from $\Delta l=1$ for $\beta_2=0$, 
while $G_{\Delta l=3}$ is smaller than $G_{\Delta l=1}$ for $\beta_2=0.15$. 
The behavior for  $\beta_2=0.15$  can be understood from the upper panels: 
for orbitals around the Fermi surface, the $\Delta l=3$ mode can have larger 
component-wise magnitudes, but its net contribution is strongly reduced by 
within-mode cancellation. In contrast, the $\Delta l=1$ mode, despite smaller 
individual terms, exhibits more constructive addition and therefore produces 
a larger net single-particle octupole energy contribution.

\subsection{Validation in Collective Rotational Spectra}

\begin{figure}[H]
    \centering
    \includegraphics[height=15cm]{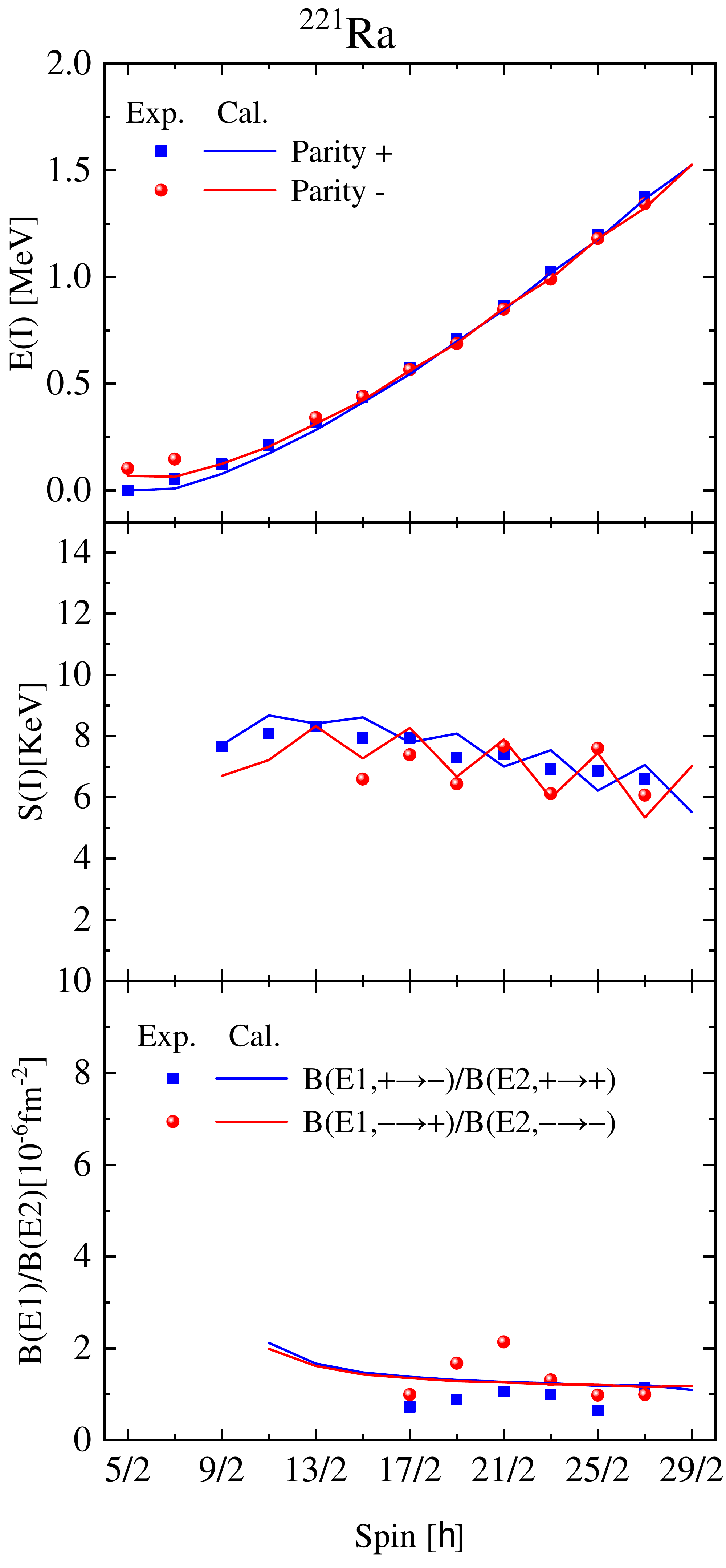}\ \ 
    \includegraphics[height=15cm]{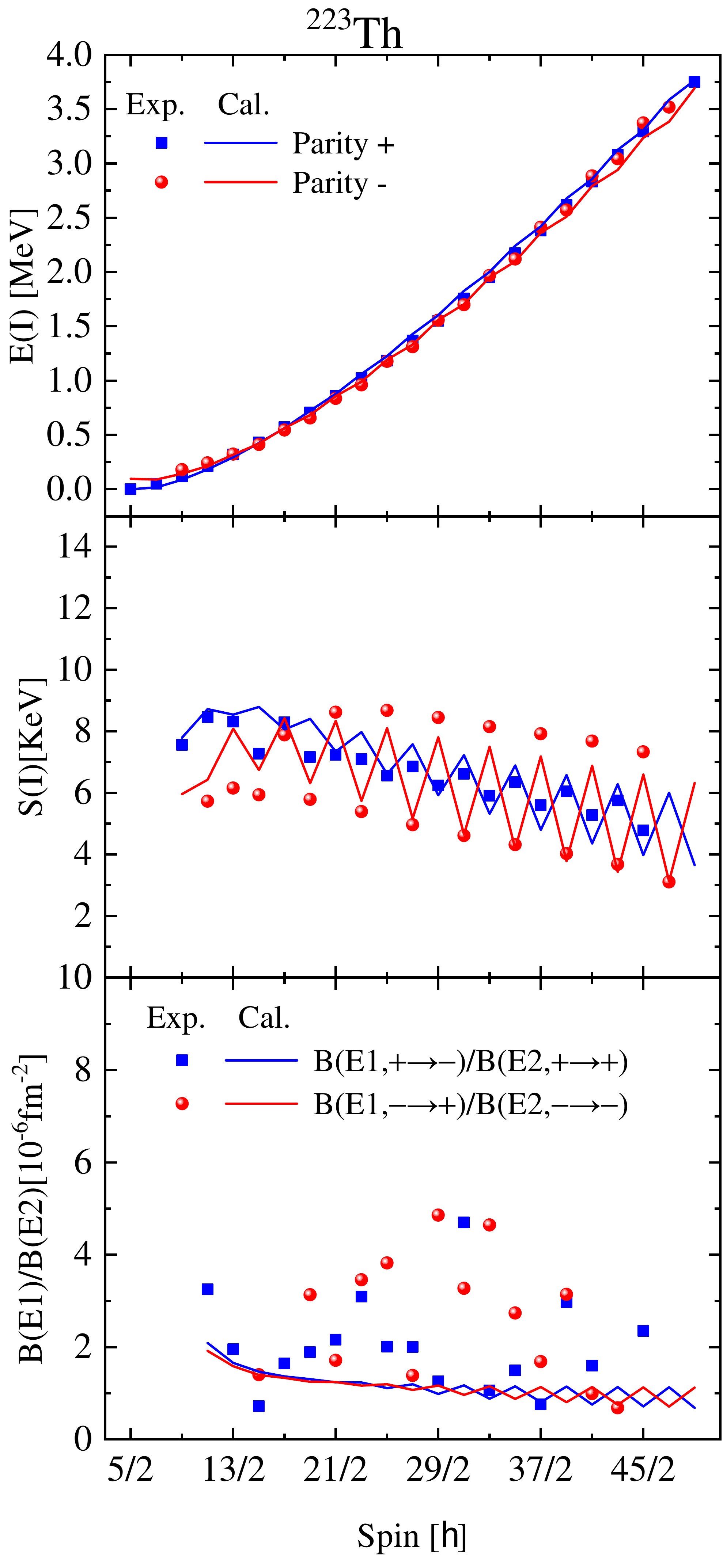}
    \caption{Comparison between the PRM results (lines) and the
    experimental data (markers)~\cite{Jain07} for $^{221}$Ra (left panel) and
    $^{223}$Th (right panel). The upper, middle and lower panels show the excitation
    energies $E(I)$, the staggering parameter $S(I)=[E(I)-E(I-1)]/(2I)$, and  the ratios
    $B(E1)/B(E2)$, respectively. Blue
    (red) denotes the positive- (negative-) parity band.}
    \label{fig:spectrum}
\end{figure}

The conclusion drawn at the single-particle level needs to be tested within
the context of real collective nuclear motion. We employ the
PRM to perform a
global fit of the level structure and electromagnetic transitions for the
typical octupole-deformed nucleus $^{221}$Ra and $^{223}$Th. As shown in
Fig.~\ref{fig:spectrum}, the theoretical calculations reproduce well the
experimentally observed low-lying excitation spectra as well as the $B(E1)$
and $B(E2)$ transition probabilities.

Fig.~\ref{fig:spectrum} compares the PRM results with the available 
data~\cite{Jain07} for the excitation energies $E(I)$, the staggering 
parameter $S(I)=[E(I)-E(I-1)]/(2I)$, and the $B(E1)/B(E2)$ ratios. The 
calculated spectra and staggering patterns for both the positive-parity band 
and the negative-parity band are reproduced well. In addition, the calculated 
$B(E1)/B(E2)$ ratios follow the experimental trends, indicating that the 
fitted PRM wave functions capture the essential interplay between collective 
rotation and reflection-asymmetric correlations in $^{221}$Ra and $^{223}$Th. 
It should be noted taht the intrinsic  moment $Q_{10}^{'}$ and  $Q_{20}^{'}$ 
are adopted by the values of empirical equations rather than the free 
parameters.

\begin{figure}[H]
    \centering
    \includegraphics[height=11cm]{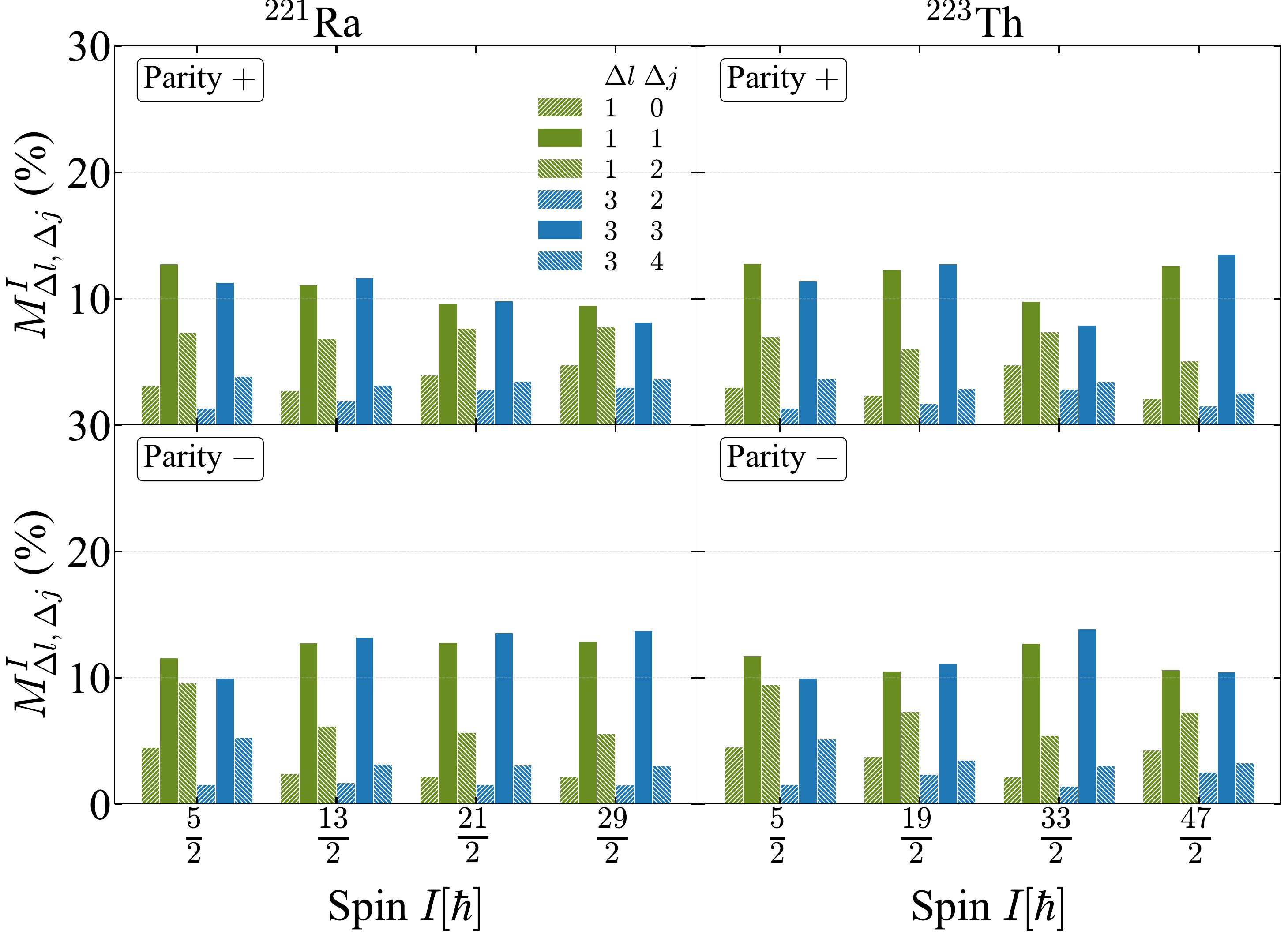}
    \caption{Mixing ratios extracted from the PRM collective wave
    functions of $^{221}$Ra (left) and $^{223}$Th (right) as functions of spin
    $I$. The calculations employ the same parameter
    set as in Fig.~\ref{fig:spectrum}. The figure displays the
    component-resolved ratios $M_{\Delta l,\Delta j}$, where green (blue)
    bars represent the $\Delta l=1$ ($\Delta l=3$) contributions and
    different hatching patterns indicate different $\Delta j$ values. The
    upper panels correspond to the positive-parity bands, whereas the lower
    panels correspond to the negative-parity bands.}
    \label{fig:collective_wf}
\end{figure}

To directly examine the presence of $\Delta l=1$ components in the collective 
states, we project the fitted PRM collective rotational wave functions onto 
the coupled basis $\ket{IMK}\ket{nlj\Omega}$ and analyze the contributions of 
different single-particle components to the collective states. 
Fig.~\ref{fig:collective_wf} summarizes the extracted mixing ratios for 
different spins $I$ in the positive-parity and negative-parity bands. The 
figure displays the component-resolved mixing ratios $M_{\Delta l,\Delta j}$, 
with the upper panels corresponding to the positive-parity band and the lower 
panels to the negative-parity band. The analysis shows that orbitals linked 
by $\Delta l=1$ couplings still contribute a significant proportion within 
the collective wave functions which successfully describe the experimental 
data. This confirms, from the perspective of collective motion,  $\Delta l=1$ 
octupole coupling constitute a substantial physical component that must be 
considered in the structure of real octupole-deformed atomic nuclei.

The key points of the multi-angle analysis  are to explicitly track which 
spherical components contribute to the mixing and to identify whether the 
dominant octupole-induced admixtures originate from $\Delta l=1$ or $\Delta 
l=3$ couplings. Based on these analysis, octupole correlations between 
single-particle orbitals with $\Delta l=1$  play a significant role at both 
the single-particle level and in collective rotational structures.

\section{Summary}

In this work, we have clarified the $\Delta l=1$ coupling in 
octupole-deformed nuclei by combining a reflection-asymmetric Nilsson 
description of intrinsic single-particle structure with particle-rotor model 
calculations of collective spectra. 

We introduce the channel-resolved mixing ratios $M_{\Delta l,\Delta j}$ and 
the component-resolved energy contributions $G_{\Delta l,\Delta j}$ to 
analyze the octupole-induced admixtures originate from $\Delta l=1$ and 
$\Delta l=3$ couplings. It is found that the leading contributions are 
typically $(\Delta l,\Delta j)=(3,3)$ and $(1,1)$. The level-averaged total 
octupole-induced mixing reaches about 30\%, and is relatively insensitive to 
quadrupole shape, with the $\Delta l=1$ component comparable to the 
traditionally emphasized $\Delta l=3$ component.  For the rotational 
structure, the calculations reproduce the main rotational features of 
$^{221}$Ra and $^{223}$Th, and the fitted collective wave functions contain 
substantial $\Delta l=1$ components.

Our central conclusion is that octupole collectivity cannot be understood 
solely within the conventional $\Delta l=3$ picture (such as 
$g_{9/2}\leftrightarrow j_{15/2}$, $f_{7/2}\leftrightarrow 
i_{13/2}$,$d_{5/2}\leftrightarrow h_{11/2}$  and $p_{3/2}\leftrightarrow 
g_{9/2}$), and $\Delta l=1$ correlations (such as $i_{11/2}\leftrightarrow 
j_{15/2}$, $h_{9/2}\leftrightarrow i_{13/2}$,$g_{7/2}\leftrightarrow 
h_{11/2}$  and $f_{5/2}\leftrightarrow g_{9/2}$) are generally 
non-negligible.  These results indicate that $\Delta l=1$ and $\Delta l=3$ 
couplings act cooperatively in driving reflection asymmetry.

%\red{Although the present numerical benchmark is centered on the
%octupole-favorable $N\approx 134$ region, the mechanism identified here is
%generic and should also apply to other octupole-magic regions
%(e.g., around $N(Z)\sim 34, 56, 88,$ and $134$), where
%deformation-renormalized near-Fermi structures allow
%$\Delta l=1$ and $\Delta l=3$ opposite-parity couplings to act jointly.}

\begin{acknowledgments}
This work is partly supported by the National Natural Science Foundation of
China (No. 12475123, No. 12225504, No. 12321005).
\end{acknowledgments}

\end{document}